\documentclass[aip,jcp,reprint,floatfix,onecolumn]{revtex4-1}
\usepackage{mathtools}
\usepackage{graphicx}              
\usepackage{bm}                    
\usepackage{amsmath}
\usepackage{tikz}
\usepackage{relsize}

\usepackage{blkarray}
\usepackage{multirow}

\def\beq{\begin{equation}}
\def\eeq{\end{equation}}
\def\det{\text{det}}
\def\tr{\text{tr}}

\def\P{R}

\def\DI{\frac{\Phi}{D}}
\def\DI{\chi}

\def\T{T}
\begin{document}

\title{Optimizing the energy with quantum Monte Carlo: A lower numerical scaling for Jastrow-Slater expansions}

\author{Roland Assaraf}
\email{assaraf@lct.jussieu.fr}
\affiliation{Sorbonne Universit\'es, UPMC Univ Paris 06, CNRS, Laboratoire de Chimie Th\'eorique (UMR7616), 4 place Jussieu F-75252 Paris, France}
\author{Saverio Moroni}
\email{moroni@democritos.it}
\affiliation{CNR-IOM DEMOCRITOS, Istituto Officina dei Materiali, and SISSA Scuola Internazionale Superiore di Studi Avanzati, Via Bonomea 265, I-34136 Trieste, Italy} 
\author{Claudia Filippi}
\email{c.filippi@utwente.nl}
\affiliation{MESA+ Institute for Nanotechnology, University of Twente, P.O. Box 217, 7500 AE Enschede, The Netherlands}

\begin{abstract}
We present an improved formalism for quantum Monte Carlo calculations of energy derivatives 
and properties (e.g. the interatomic forces),  with a multideterminant Jastrow-Slater  function.  
As a function of the number  $N_e$ of Slater determinants, the numerical scaling of $O(N_e)$ per 
derivative we have recently reported is here lowered to $O(N_e)$ for the entire set of derivatives.  
As a function of the number of electrons $N$, the scaling to optimize the wave function and the 
geometry of a molecular system is lowered to $O(N^3)+O(N N_e)$, the same as computing the energy 
alone in the sampling process. The scaling is demonstrated on linear polyenes up to C$_{60}$H$_{62}$ 
and the efficiency of the method is illustrated with the structural optimization of butadiene and 
octatetraene with Jastrow-Slater wave functions comprising as many as 200000 determinants and 
60000 parameters.
\end{abstract}
\maketitle

\section{Introduction}

Quantum Monte Carlo methods (QMC) are first-principle methods which can efficiently solve
the Schr\"odinger equation.  For fermionic systems, they  are powerful variational approaches
because they can handle a large variety of variational wave functions $\Psi({\bf R})$, where
${\bf R}=({\bf r}_1 \dots {\bf r}_N)$ represents the coordinates of the $N$ electrons of the
system.  Here, the vector ${\bf r}_i =(x_i,y_i,z_i,\sigma_i)$ indicates the 3 spatial
coordinates of the electron $i$, $(x_i,y_i,z_i)$ and its spin component $\sigma_i$ ($\sigma_i
=\pm \frac{1}{2}$).  This flexibility stems from the fact that  integrals are not computed
analytically but from a stochastic sampling.  For example, the variational energy is
\begin{equation}
E = 
{\int  d{\bf R} \Psi^2({\bf R}) \frac{\hat{H}\Psi}{\Psi}({\bf R}) }\,
\end{equation}
where $\hat{H}$ is the Hamiltonian and $\Psi$ is normalized, and can be interpreted as the
expectation value of a random variable, the so-called local energy $E_L = {\hat{H}\Psi}/{\Psi}$
on the probability density $\Psi^2({\bf R})$.  QMC methods can be used as benchmark methods also
for relatively large systems thanks to their favorable scaling with the number of particles $N$.  
For a given parametrization of  $\Psi$,  $E$ is typically computed with a scaling $O(N^2)$ in 
memory requirements and $O(N^3)$ in CPU per Monte Carlo step.  In practice,  one needs to optimize 
the parameters of $\Psi$ and the geometry of a molecular system.  Despite the availability of 
stable wave function optimization methods \cite{Umrigar07}, such techniques remain costly and
one of the main reasons is that a large number of derivatives of $E$ (typically $O(N^2)$) 
has to be computed.  Lowering the numerical scaling per derivative is therefore important.
For single determinants,  Sorella {\sl et  al.} have found that the low-variance estimators 
of the $3N_{\rm atoms}=O(N)$ intermolecular forces can be calculated with a scaling $O(N^3)$
instead of $O(N^4)$ with the use of algorithmic differentiation techniques \cite{Sorella10}. 
We have recently recovered the  same reduction using transparent matricial formulas and extended 
it to the $O(N^2)$ orbital coefficients \cite{Filippi_ci2016}. For expansions over additional 
$N_e$ Slater determinants, $D_i$, multiplied by a positive Jastrow correlation factor $J ({\bf R})$,
\begin{equation}
\Psi ({\bf R}) = J ({\bf R}) \Phi({\bf R}) =  J ({\bf R})  \sum_{i=0}^{N_e} c_i D_i \,,
\label{cijas}
\end{equation}
Clark {\sl et al.} have proposed a method to compute $\Psi$ with a scaling $O(N_e)$ 
and $E_L$ with a scaling  $O(N N_e)$~\cite{clark11} that we have further reduced 
to $O(N_e)$ and extended to any derivative of $E_L$ \cite{Filippi_ci2016}. The derivatives 
of $E_L$ are useful because they are involved in low-variance estimators for forces and observables
\cite{zero_variance,filippi00,assaraf03}.  At the origin of this reduction is the observation
that the local energy can be written in terms of a first-order (logarithmic) derivative of
the determinantal component, $\partial_\lambda \Phi/\Phi$.

In this paper, we show that the scaling $O(N_e)$ per derivative can be further improved to 
$O(N_e)$ for {\it any} set of derivatives of  $\Psi$ and $E_L$. The core observation is that the determinantal 
part $\Phi$ is a function of the matrix elements $\tilde{A}_{ij} = \phi_j ({\bf r}_i)$ where $\phi_j$ 
is an orbital and $i$ an electron index, and that any derivative of $\Phi$ can be computed using 
a simple trace formula involving the matrix $\Gamma$ defined as the logarithmic gradient of $\Phi$ 
with respect to $\tilde{A}$.  
The first derivatives of the local energy $\partial_\mu E_L$ can then be expressed as traces involving 
$\Gamma$ and {\it one} of its derivative $\partial_\lambda \Gamma$:  many derivatives of $\Psi$ 
and $E_L$ are obtained efficiently because the matrices $\Gamma$ and $\partial_\lambda \Gamma$ 
are computed only once for the whole set of parameters $\{\mu\}$. Consequently, 
the calculation of all derivatives of $E$ with 
respect to all parameters of the wave function (Jastrow parameters, orbital coefficients,
the coefficients of the expansion $\{c_i\}$, and all nuclear positions) has now the same
scaling as the calculation of $E$ alone, opening the path to full optimization of large
multideterminant expansions.

In the next Section, we outline the main idea and introduce the matrix $\Gamma$. In 
Section~\ref{sec3}, we present a formula  to compute $\Gamma$ at a cost $O(N^3)+O(N_e)$ 
and, in Section~\ref{sec4}, discuss the formulas for the second derivative of $\Phi$ and, 
specifically, the first derivatives of $E_L$.  In Section~\ref{sec5}, we demonstrate 
the scaling of the computation of interatomic forces with multideterminant wave functions 
on polyenes up to C$_{60}$H$_{62}$ and, in the last Section, apply the scheme to the 
optimization of multideterminant wave functions and geometries of butadiene and octatetraene.

\section{Derivative of the determinantal expansion}
\label{sec2}

The determinantal component $\Phi$ in the Jastrow-Slater expansion of Eq.~(\ref{cijas}) is a 
linear combination of $N_e+1$ Slater determinants 
\begin{equation}
\Phi =  \sum_{I=0}^{N_e} c_I \det ({A}_I) \,.
\label{phidef}
\end{equation}
For a system including $N$ electrons, the matrix $A_I$ is an $N\times N$ Slater matrix, built from  
$N$ of the $N_{\rm orb}$ molecular spin-orbitals  $\phi_i({\bf r})$ $( 1 \leq i \leq N_{\rm orb})$. Mathematically, 
$A_I$ comprises $N$ columns of the $N \times N_{\rm orb}$  matrix $\tilde{A}$ defined as follows
\begin{equation}
\tilde{A}_{ij} = \phi_j ({\bf r}_i)\,.
\end{equation}

In general, one needs to compute many derivatives of $\Phi$ with respect to different parameters 
of $\tilde{A}$.  These parameters can be the electron coordinates, nuclei coordinates, orbital 
coefficients, basis-function parameters and so on. The derivative of $\Phi$ with respect to a 
given parameter $\mu$ in $\tilde{A}$ is obtained from the chain rule
\begin{equation}
\partial_\mu \ln( \Phi) = \frac{\partial \ln (\Phi)}{\partial \tilde{A}_{ij}} \partial_\mu {\tilde{A}}_{ij} 
                        = \tr(\Gamma \partial_\mu \tilde{A})\,,
\label{gentrace1}
\end{equation}
where a summation on repeated indices is implied and we have introduced 
$\Gamma$, that is,  the gradient of $\ln(\Phi)$  with respect to the matrix elements of $\tilde{A}$
\begin{equation}
\Gamma_{ji} = \frac{\partial \ln \Phi}{\partial \tilde{A}_{ij}}\,.
\label{gendiff}
\end{equation}
The trace formula (\ref{gentrace1}) is at the core of greater efficiency in computing many derivatives 
of $\Phi$ because the $N\times N_{\rm orb}$ matrix $\Gamma$ depends only on $\tilde{A}$ and not on 
$\partial_\mu \tilde{A}$.  For a given configuration ${\bf R}$ in the Monte Carlo sample, $\Gamma$ is 
computed only once for all the set of derivatives.  In addition, $\Gamma$  can be evaluated efficiently, 
at a cost $O(N^3)+O(N_e)$ as we will see in the next Section.  Once $\Gamma$ is computed and stored, any 
new derivative $\partial_\mu \ln( \Phi)$ requires to calculate besides $\partial_\mu \tilde{A}$ the trace 
(\ref{gentrace1}) at a cost $O(N_{\rm orb} \times N)$. What is important here is that this scaling is 
independent on $N_e$ and leads to vast improvements over previous methods \cite{Filippi_ci2016, clark11}
 when $N_e$ and the number of derivatives are large.

Finally, also quantities like the local energy  or  the value of the wave function after one electron move, 
can be computed using this  trace formula  (\ref{gentrace1}). This is because one-body operators can be 
also expressed as first order derivatives of $\ln \Phi$  when applied to a Jastrow-Slater expansion  
\cite{Filippi_ci2016}.

\section{Efficient evaluation of the matrix $\Gamma$}
\label{sec3}

\subsection{Convenient expression for $\Phi$}

The determinants of the Slater matrices $A_I$ can be computed efficiently because $A_I$ usually differs 
by a few columns from a reference Slater  matrix $A$.  For example, let $A$ be the $4\times 4$
Slater matrix built with the orbitals $\phi_1, \phi_2,\phi_3,\phi_4$:
\begin{eqnarray}
A  & =& \left( \begin{array}{ccccc} \tilde{A}_1 &  \tilde{A}_2 & \tilde{A}_3 & \tilde{A}_4
\end{array} \right) \,,
\end{eqnarray}
where the notation $\tilde{A}_i$ stands for the $i^{th}$ column of $\tilde{A}$.
The Slater matrix of a double excitation $(3, 4) \to (5, 7)$ is
\begin{eqnarray}
A_I & = & \left( \begin{array}{ccccc} \tilde{A}_1 &  \tilde{A}_2  & \tilde{A}_5 & \tilde{A}_7 
\end{array} \right) \,.
\end{eqnarray}
Here, $A_I$ and $A$ differ only in the 2 last columns.  The determinant of $A_I$ is 
\begin{equation}
\det(A_I)  =  \det (A) \det (A^{-1} A_I) \nonumber
\end{equation}
and
\begin{eqnarray}
A^{-1}A_I  & = &    \left( \begin{array}{ccccc}
A^{-1}\tilde{A}_1  & A^{-1}\tilde{A}_2  & A^{-1}\tilde{A_5} & \tilde{A}^{-1} \tilde{A}_7 \end{array}\right) 
 =  
\left( \begin{array}{cccc}
1 \, \, & 0 \, \, &  (A^{-1}\tilde{A})_{15}   & (A^{-1}\tilde{A})_{17}   \\ 
0 \, \,& 1 \, \, & (A^{-1}\tilde{A})_{25}  &  (A^{-1}\tilde{A})_{27} \\ 
0 \, \, & 0 \, \,&  (A^{-1}\tilde{A})_{35} & (A^{-1}\tilde{A})_{37}    \\
0 \, \, & 0 \, \, &   (A^{-1}\tilde{A})_{45} & (A^{-1}\tilde{A})_{47}   \\
\end{array} \right) \,,
\end{eqnarray}
where a column of the identity matrix arises whenever $A_I$ and $A$ share the
same column.  The determinant of $A^{-1} A_I$ is readily evaluated:
\begin{equation}
\det (A^{-1}A_I) = \det \left(  \begin{array}{cc}
   (A^{-1}\tilde{A})_{35} & (A^{-1}\tilde{A})_{37}    \\
  (A^{-1}\tilde{A})_{45} & (A^{-1}\tilde{A})_{47}   \\
\end{array} \right)\,. \label{alpha_I}
\end{equation}
More generally, the determinant of $A^{-1}A_I$ for a $k_I^{th}$-order excitation is the 
determinant of a $k_I \times k_I$ submatrix.
Such a submatrix can always be written  as follows
\begin{equation}
\alpha_I = P_I^T A^{-1}\tilde{A} Q_I\,,
\end{equation}
where, in our example,
\begin{equation}
P_I = 
\left( \begin{array}{cccccc}
0 & 0  \\ 
0 & 0  \\
1 & 0  \\
0 & 1  \\
\end{array} \right)
\label{expi}
\end{equation}
and 
\begin{equation}
Q_I = \left( \begin{array}{cccccc}
0 & 0  \\ 
0 & 0  \\ 
0 & 0  \\
0 & 0  \\
1 & 0  \\
0 & 0  \\
0 & 1  \\
0 & 0 \\
\vdots & \vdots \\
\end{array} 
\right)\,.
\label{exqi}
\end{equation}
In general, $P_I$ is  such that $A P_I$ are the columns of $A$ which differ from those of $A_I$, and  
$Q_I$  is such that $\tilde{A}Q_I=A_I P_I$. In other words  $P_I$ (applied on the right of $A$) selects 
the columns of $A$ from which excitations are built, and $Q_I$ (applied on the right of $\tilde{A}$) 
selects the columns of $\tilde{A}$ to which excitations are built.  To summarize, the expression 
\begin{equation}
\det(A_I) = \det(A) \det (P_I^T A^{-1}\tilde{A}Q_I) 
\label{A_I}
\end{equation}
enables to compute the determinant of a large $N\times N$ matrix as the determinant of a small $k_I \times k_I$ 
submatrix  of $A^{-1}\tilde{A}$.  This expression can also be proven using the determinant 
lemma  \cite{clark11,Filippi_ci2016}.  Finally, the convenient expression for $\Phi$ to efficiently compute $\Gamma$ is:
\begin{eqnarray}
\Phi & = & \det(A) \times  \sum_I c_I \det (P_I^T A^{-1}\tilde{A}Q_I)  \,.
\label{phiprod}
\end{eqnarray}

\subsection{Convenient expression for $\Gamma$}

Introducing the matrix $\P$ such that $A= \tilde{A}\P$, the expression (\ref{phiprod}) is explicitly a function 
of $\tilde{A}$.  In particular, the summation on the r.h.s. of Eq.~(\ref{phiprod})
\begin{equation}
\DI \equiv \frac{\Phi}{\det(A)} = \sum_I c_I \det (P_I^T A^{-1}\tilde{A}Q_I)
\label{chi}
\end{equation}
is a polynomial function depending on the matrix elements of 
\begin{equation}
\T \equiv A^{-1}\tilde{A} =  (\tilde{A}R)^{-1} \tilde{A} \,.
\end{equation}
The order of this polynomial is the order of the highest-order exitation. It is usually low (typically $k_I < 4$).
Applying the chain rule and using the convention of summation over repeated indices, we obtain
\begin{eqnarray}
\partial_\mu \ln (\Phi) &=& \partial_\mu \ln \det(A) +  \partial_\mu \ln \DI \nonumber \\
  & =& \tr(A^{-1}  \partial_\mu A) + 
  \frac{\partial \ln \DI}{\partial \T_{ij}} \partial_\mu \T_{ij} \nonumber
\\ 
 & = &  \tr(A^{-1}  \partial_\mu A ) + \tr(Y  \partial_\mu \T  )\,,
\label{ymder}
\end{eqnarray}
where 
\begin{equation}
  Y_{ji} \equiv   \frac{\partial \ln \DI}{\partial \T_{ij}}= \frac{1}{\DI}\frac{\partial \DI}{\partial \T_{ij}}\,.
\end{equation}
It is simple to show that
\begin{eqnarray}
\frac{\partial \DI}{\partial \T_{ij}}=\sum_{I>0}^{N_e}c_I\det(\alpha_I)(Q_I\alpha_I^{-1}P_I^T)_{ji}\,.
\label{eqy}
\end{eqnarray}
The derivative of $\T$ is given by
\begin{equation}
\partial_\mu \T  =  -A^{-1}   \partial_\mu A  \, \,  A^{-1}\tilde{A} + A^{-1} \partial_\mu \tilde{A}\,.
\label{dchi}
\end{equation}
Finally, writing $A =  \tilde{A} \P$ and using the cyclic property of the trace, we obtain 
\begin{equation}
\partial_\mu \ln(\Phi) = \tr( \Gamma \partial_\mu \tilde{A})\,,
\end{equation}
where 
\begin{eqnarray}
\Gamma  & = & \P A^{-1} +\ (1-\P A^{-1}\tilde{A}) Y A^{-1} \nonumber \\
& = & \left[ \P(1-A^{-1}\tilde{A}Y)+Y\right] A^{-1}\,.
\label{gammeff}
\end{eqnarray}
For example, if the occupied orbitals are the $N$ first ones, the matrix $\Gamma$ is 
\begin{equation}
\Gamma  =  \left( \begin{array}{c}
   A^{-1}-A^{-1}\tilde{A}YA^{-1}      \\ 
 Y_{{\rm virt}} A^{-1} 
\end{array} \right)\,,
\end{equation}
where the first line is a $N\times N$ matrix. The second line is a $(N_{\rm orb}-N)\times N$ matrix 
where $Y_{{\rm virt}}$ represents the non-zero lines of $Y$, i.e. the last  $N_{\rm virt} \equiv N_{\rm orb}-N$ lines.

\subsection{One-body operators and first-order derivatives of $\Phi$}

First-order derivatives of $\Phi$ can be computed with the trace formula (\ref{gentrace1}) which involves the $\Gamma$
matrix.  One-body operators acting on the wave function can be also expressed as first-order derivatives of $\ln \Phi$  
when applied to a Jastrow-Slater expansion as we have shown in Ref.~\onlinecite{Filippi_ci2016}. The local energy 
for example can be written as a first-order logarithmic derivative of the determinantal part where $\tilde{A}$ has 
been replaced by
\begin{eqnarray}
\tilde{A}_\lambda & =  & \tilde{A}+ \lambda \tilde{B}
\end{eqnarray}
and $\tilde{B}$ is an appropriate matrix depending on the orbitals, the Jastrow factor, and their derivatives.
In particular, the reference Slater determinant $A$ has been replaced by $A_\lambda = A + \lambda B$.
The determinantal part of the wave function is now 
\begin{equation}
\Phi= \det(A_\lambda) \left[\sum_I c_I \det (P_I^T A_\lambda^{-1} \tilde{A}_\lambda Q_I)\right]\,.
\end{equation}
From this expression, one can compute the local energy 
\begin{eqnarray}
E_L & = &  \partial_\lambda (\ln \Phi) 
=  \tr(\Gamma \tilde{B})\,.
\label{E_L}
\end{eqnarray}
In the presence of the Jastrow factor, one recovers the same trace expression for the local energy 
of $\Psi$ but with a matrix $\tilde{B}$ also depending on $J ({\bf R})$ and its derivatives \cite{Filippi_ci2016}.

\section{Second-order derivatives}
\label{sec4}

The second derivative of $\Phi$ can be written in terms of $\Gamma$ and its derivative as
\begin{eqnarray}
\partial_{\lambda} \partial_\mu \ln (\Phi) &= &\partial_\lambda  \tr(\Gamma  \partial_\mu \tilde{A}) \nonumber \\
 & =&  \tr(\Gamma \, \partial_{\lambda \mu} \tilde{A}) +  \tr( \partial_\lambda \Gamma \, \, \partial_\mu \tilde{A}) \,.
\label{d_lambdaE}
\end{eqnarray}

\subsection*{Example of the derivative of the local energy}

When computing improved estimators of derivatives of the energy $E$, we need also the derivatives of the local 
energy $E_L$. It follows from Eq.~\ref{E_L} that  
the derivative of the local energy with respect to a given parameter $\mu$ is 
\begin{eqnarray}
\partial_\mu E_L &=& \partial_{\lambda} \partial_\mu \ln (\Phi) \nonumber \\
 & =&  \tr(\Gamma \, \partial_{\mu} \tilde{B}) +  \tr( \partial_\lambda \Gamma \, \, \partial_\mu \tilde{A})\,.
\label{d_lambdaE}
\end{eqnarray}
The order of the derivation has been chosen so that $\tilde{A}$ and not $\Gamma$ is differentiated with respect to 
$\mu$. Consequently, the matrix $\partial_\lambda \Gamma$ does not depend on the  parameter $\mu$ and has to be computed  
only once, whatever the number of second derivatives we need.  Once $\partial_\lambda \Gamma$ has been computed, the 
calculation of $\partial_\mu E_L$ involves (besides $\partial_\mu \tilde{A}$ and $\partial_{\mu} \tilde{B}$)
two traces which can be computed at a cost $O(N N_{\rm orb})$.
Importantly, such a calculation does not depend on $N_e$ contrary to what was presented in   Ref.~\onlinecite{Filippi_ci2016}.

\subsection*{Efficient calculation of $\partial_\lambda \Gamma$}

The derivative of $\Gamma$ is
\begin{eqnarray}
 \partial_\lambda \Gamma  & =&  
 \left[-\Gamma  B  +\partial_\lambda Y
 + R( \partial_\lambda \T \, \, Y +\T \partial_\lambda Y)\right]A^{-1}\,,
\label{gammader}
\end{eqnarray}
where 
\begin{eqnarray}
\partial_\lambda \T
  & = & A^{-1} (\tilde{B} - B\T)\equiv \tilde{M}\,.
\label{tildeMdef}
\end{eqnarray}
Applying the chain rule, we obtain
\begin{eqnarray}
\partial_\lambda Y_{ij} 
 & =& 
Z_{ijkl} \tilde{M}_{kl}\,, 
\end{eqnarray}
where 
\begin{eqnarray}
Z_{ijkl} &\equiv& \frac{\partial^2 \ln \DI}{\partial \T_{ij} \partial \T_{kl}} \\
& = &  \frac{1}{\DI} \frac{\partial^2 \DI}{\partial \T_{ij} \partial \T_{kl}}- Y_{ij}Y_{kl}\,.
\end{eqnarray}
It follows from Eq.~\ref{eqy} that
\begin{eqnarray}
\frac{\partial^2 \DI}{\partial \T_{ij} \partial \T_{kl}}
 =\sum_{I>0}^{N_e}c_I\det(\alpha_I)\left[(Q_I\alpha_I^{-1}P_I^T)_{ji} (Q_I\alpha_I^{-1}P_I^T)_{lk}
         -(Q_I\alpha_I^{-1}P_I^T)_{jk} (Q_I\alpha_I^{-1}P_I^T)_{li}\right]\,.
\label{d2chi}
\end{eqnarray}
We can compute the derivatives of $\DI$ avoiding the evaluation of inverse matrices. That  will be presented 
in the appendix. 

\subsection*{Derivatives with respect to the linear coefficients}

The derivatives of a local quantity with respect to the expansion coefficients require instead to evaluate
the action of the one-body operator on each excited determinant $A_I$ separately (Eq.~\ref{A_I}). For instance, 
as we have shown in Ref.~\onlinecite{Filippi_ci2016}, the derivative of the local energy with respect to $c_I$ is given by
\begin{eqnarray}
\partial_{c_I}E_L &=& \partial_\lambda\partial_{c_I} (\ln\Phi)\nonumber\\
&=&\frac{\det(A)}{\Phi}\partial_\lambda \det(P_I^T T Q_I)\nonumber\\
&=&\frac{\det(A)}{\Phi}\tr(\alpha_I^{-1}P_I^T\partial_\lambda T Q_I)\nonumber\\
&=&\frac{\det(A)}{\Phi}\tr(\alpha_I^{-1}P_I^T\tilde{M}Q_I)\,.
\end{eqnarray}
These quantities are needed in the optimization of the energy with respect to the linear coefficients 
and can be computed at a cost $O(N_e)$.

\section{Numerical scaling}
\label{sec5}

In practice, for each step of the Monte Carlo algorithm, we need to compute $\tilde{A}$, $A^{-1}$, and 
$T=A^{-1}\tilde{A}$ at a cost of at most $O(N^3)$ (products and inversions of matrices).  Then, we need to 
calculate the first and second derivatives of $\chi$ with respect to $T$ (Eqs.~\ref{eqy} and \ref{d2chi})
at a cost $O(N_e)$ (a few sums and products for each excitation). The related tensors $Y$ and $Z$ are also 
computed at a cost $O(N_e)$. $\partial_\lambda Y$ is computed at a cost $O(N_d)$ where $N_d$ is the total 
number of double excitations involved in any $k_I^{th}$ order excitation ($k_I \geq 2$), where of course 
$N_d< N_e$.  
Finally, $\Gamma$ and $\partial_\lambda \Gamma$ are computed at a cost $O(N^3)$ (product of matrices).

In particular, computing the $3N_{\rm atoms}$ components of the inter-atomic forces with improved estimators has 
a scaling 
\begin{equation}
O (N^3)+ O(N^2 N_{\rm atoms})+ O(N_{\rm atoms} N N_{\rm virt}) + O(N_e)\,.
\end{equation}
Assuming that $N_{\rm virt}=O(N)=O(N_{\rm atoms})$, this scaling simplifies
\begin{equation}
O(N^3) + O(N_e)\,.
\end{equation}
This is significantly more efficient  than the scaling \footnote{or  $O(N^3) + O(N^2 N_{\rm act} N_{\rm atoms}) + 
O(N_e N_{\rm atoms})$} $O(N^3) + O(N^2 N_{\rm virt} N_{\rm atoms}) + O(N_e N_{\rm atoms})$ presented in our previous 
work~\cite{Filippi_ci2016}, in the large $N_e$, $N_{\rm atoms}$  or $N$, $N_{\rm atoms}$, $N_{\rm virt}$ regimes.
The term $O(N^2 N_{\rm act} N_{\rm atoms})$ is no more present because here we avoid to compute $\partial_\mu \T$.
Regarding the sampling process, when one-electron moves are used (see appendix), the total numerical cost for a 
full sweep (all the electrons are moved once) is  $\sim O (N^3)+ O(N N_e)$.

\begin{figure}[h]
\includegraphics[width=\columnwidth]{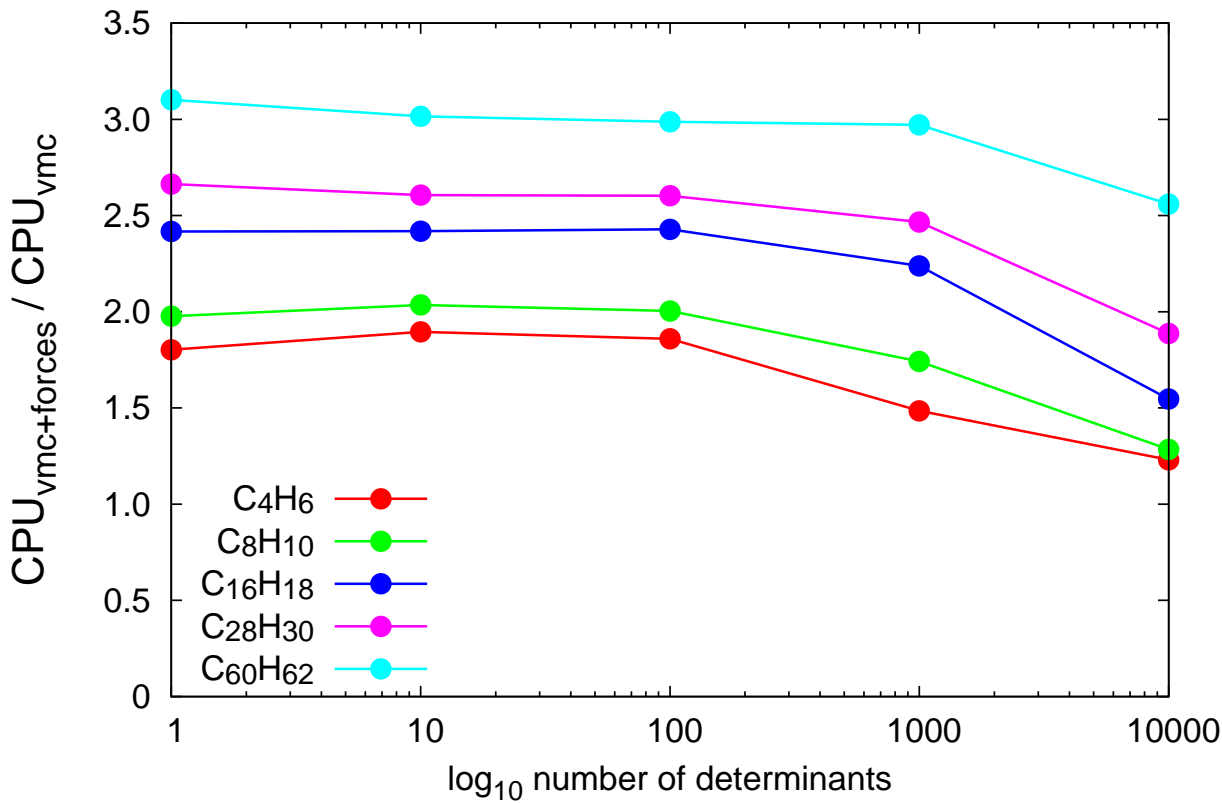}
\caption{
Ratio of the CPU time for a VMC calculation of the forces to the CPU time for the same 
simulation of the energy alone for the sequence of molecules C$_n$H$_{n+2}$ with $n$ between 4 
and 60 and an increasing number of determinants in the Jastrow-Slater wave function. The forces 
are calculated after moving all the electrons once.  
}
\label{forces_a}
\end{figure}

In Fig.~\ref{forces_a}, we demonstrate this favorable scaling in the variational Monte Carlo (VMC) computation of 
the interatomic forces for multi-determinant Jastrow-Slater wave functions using the sequence of molecules C$_n$H$_{n+2}$ 
with $n$ between 4 and 60.  For each system, the ratio of the CPU time of computing all interatomic forces to the time of 
evaluating only the energy is initially constant and then decreases when the number of determinants exceeds about 100.  
For the largest C$_{60}$H$_{62}$, computing all interatomic gradients costs less than about 3 times a VMC simulation 
where one only evaluates the total energy.
Finally, as it is shown in the Appendix, if we move one electron, many quantities can be updated so that, for each 
Monte Carlo step, the scaling is reduced to $O(N^2)+O(N_e)$.  This leads to an overall scaling $O(N^3)+O(N_e N)$ when 
all the electrons have been moved.  For an all-electron-move algorithm, the scaling is $O(N^3) + O(N_e)$ which could be more efficient when $N_e$ is large.

\section{Numerical results}
\label{sec6}

We demonstrate the formulas above on the ground-state structural optimization in VMC of butadiene (C$_4$H$_{6}$) and 
octatetraene (C$_8$H$_{10}$) using large expansions in the determinantal component of the Jastrow-Slater wave function. 
All expansion coefficients, orbital and Jastrow parameters in the wave function are optimized together with the geometry. 
Given the large number of variational parameters (up to 58652) we employ the stochastic reconfiguration optimization
method~\cite{Rocca2007} in a conjugate gradient implementation~\cite{Neuscamman2012} which avoids building and
storing large matrices. In most of our calculations, to remove occasional spikes in the forces, we use an improved 
estimator of the forces obtained by sampling the square of a modified wave function close to the 
nodes~\cite{Attaccalite2008}. To optimize the geometry, we simply follow the direction of steepest descent and 
appropriately rescale the interatomic forces.
We employ the CHAMP code~\cite{Champ} with scalar-relativistic energy-consistent Hartree-Fock pseudopotentials and 
the corresponding cc-pVXZ~\cite{Burkatzki07,Hpseudo} and aug-cc-pVXZ~\cite{aug-basis} basis sets with X=D,T, and Q.  
The Jastrow factor includes two-body electron-electron and electron-nucleus correlation terms.  
The starting determinantal component of the Jastrow-Slater wave functions before optimization is obtained in 
multiconfiguration-self-consistent-field calculations performed with the program GAMESS(US)~\cite{Gamess1993,Gordon2011}.

\begin{figure}[h]
\includegraphics[width=\columnwidth]{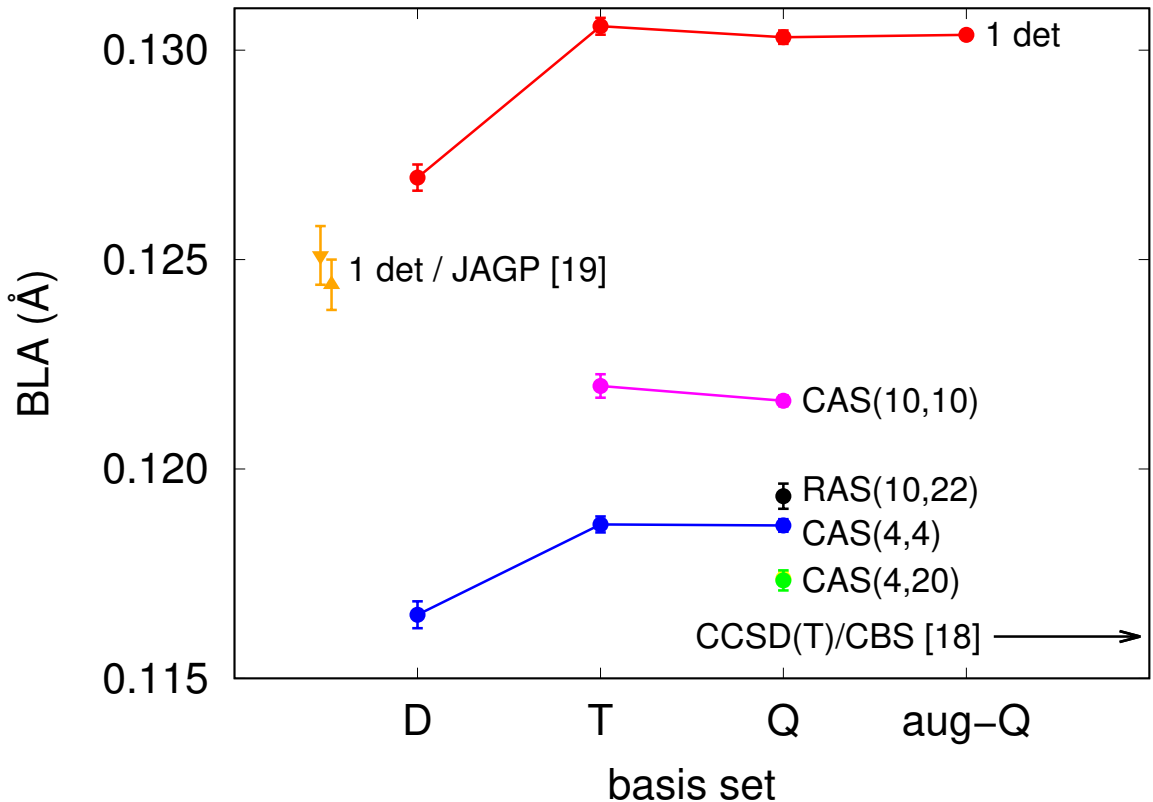}
\caption{
Bond length alternation (BLA) of C$_4$H$_6$ optimized in VMC for different basis sets and choices in the 
determinantal part of the Jastrow-Slater wave function.  The atomic positions and all parameters of the 
wave function (expansion coefficients, orbital and Jastrow parameters) are simultaneously optimized.
The CCSD(T) BLA in the CBS limit computed with various corrections~\cite{Feller2009} and the best 
value obtained with a Jastrow-antisymmetrized geminal power (JAGP)~\cite{Barborini2015a} are reported.  
}
\label{bla_buta}
\end{figure}

\begin{table}
\begin{tabular}{lcccccc}
\hline
Expansion & No.\ det & No.\ param. & C-C & C=C & BLA (\AA) \\
\hline
1 det     &  1       & 1404       & 1.45513(12) & 1.32482(05) & 0.13031(16) \\ 
CAS(4,4)  &  20      & 1547       & 1.45211(10) & 1.33347(07) & 0.11865(15) \\
CAS(4,16) &  7232    & 4995       & 1.45160(15) & 1.33422(13) & 0.11738(16) \\
CAS(4,20) &  18100   & 9147       & 1.45143(16) & 1.33409(07) & 0.11734(24) \\
CAS(10,10)&  15912   & 6890       & 1.45858(09) & 1.33694(06) & 0.12163(13) \\
RAS(10,22)&  45644   & 11094      & 1.45705(17) & 1.33760(15) & 0.11945(29) \\
CCSD(T)/CBS$^a$  &   &        & 1.4548      & 1.3377      & 0.1171      \\
CCSD(T)/CBS-corr$^b$&&        & 1.4549      & 1.3389      & 0.1160      \\
\hline
\multicolumn{6}{l}{$^a$ Ref.~\onlinecite{Feller2009}; $^b$ Ref.~\onlinecite{Feller2009}, including a CCSDT(Q)(FC)/cc-pVDZ correction.}
\end{tabular}
\caption{Optimal bond lengths and BLA values (\AA) of butadiene computed in VMC with the cc-pVQZ basis set and 
various choices of Jastrow-Slater expansions. The numbers of determinants and optimized parameters in the wave function are listed.
}
\label{table1}
\end{table}

We first focus on the VMC geometrical optimization of butadiene. Despite its small size and apparent simplicity,
predicting the bond length alternation (BLA) of butadiene remains a challenging task for quantum 
chemical approaches which lead to a spread of BLA values, mainly clustered around either 0.115 or 0.125 \AA\ (see Table 2 in 
Ref.~\onlinecite{Barborini2015a} for a recent compilation of theoretical predictions). In particular, 
Barborini and Guidoni~\cite{Barborini2015a} using
VMC in combination with Jastrow-antisymmetrized geminal power (JAGP) wave functions find a best BLA value of 0.1244(6) 
\AA, rather close to the BLA of 0.1251(7) \AA\ they obtain using a single-determinant Jastrow-Slater wave function 
and clearly distinct from the CCSD(T) prediction of 0.116 \AA\ computed in the complete basis set (CBS) limit 
and corrected for core-valence correlation, scalar-relativistic effects, and inclusion of quadruples~\cite{Feller2009}. 

To elucidate the origin of this difference, 
we consider here various expansions correlating the $\pi$ and $\sigma$ 
electrons: a) a single determinant; b) the complete-active-space CAS(4,4), CAS(4,16), and CAS(4,20) expansions (20, 7232, 
and 18100 determinants, respectively) of the four $\pi$ electrons in the bonding and antibonding $\pi$ orbitals 
constructed from the $2p_z$, $3p_z$, $3d_{xz}$, $3d_{yz}$, and $4p_z$ atomic orbitals; c) a CAS(10,10) correlating the 
six $\sigma$ and four $\pi$ electrons of the carbon atoms in the corresponding bonding and antibonding $\pi$ and 
$\sigma$ orbitals 
(15912 determinants); d) the same CAS(10,10) expansion augmented with single and double excitations in the external 
space of 12 $\pi$ orbitals and truncated with a threshold of 2$\times$10$^{-4}$ on the coefficients of the spin-adapted 
configuration state functions. This last choice results in a total of 45644 determinants and is denoted as a 
restricted-active-space RAS(10,22) expansion.

We start all runs from the same geometry and, after convergence, average the geometries over an additional 30-40 
iterations.  The results of these structural optimizations are summarized in Fig.~\ref{bla_buta}.  We find that 
the basis sets of triple- and quadruple-$\zeta$ quality yield values of BLA which are compatible within 1-1.5 
standard deviations, namely, to better than 5$\times$10$^{-4}$ \AA. The further addition of augmentation 
does not change the BLA as shown in the one-determinant case. 
In the following, we therefore focus on the cc-pVQZ bond lengths and BLA values of butadiene, which are 
summarized in Table~\ref{table1}.

With a one-determinant wave function (case a), we obtain a BLA of 0.1303(2) \AA\, which is higher than the value 
of 0.1251(6) \AA\ reported in Ref.~\onlinecite{Barborini2015a}, possibly due to their use of a basis set of quality
inferior to triple-$\zeta$. Moving beyond a single determinant, we observe a strong dependence of the result on the 
choice of active space. The inclusion of $\pi$-$\pi$ correlation within 4, 16, and 20 $\pi$ orbitals (case b) 
significantly decreases the BLA with respect to the one-determinant case with the CAS(4,16) and CAS(4,20) 
expansions yielding a BLA of 0.117 \AA\ in apparent agreement with the CCSD(T)/CBS estimate of 0.116 \AA.
Accounting also for $\sigma$-$\pi$ and $\sigma$-$\sigma$ correlations in a CAS(10,10) (case c) leads however to a more
substantial lengthening of the single than the double bond and a consequent increase of BLA. Finally, allowing 
excitations out of the CAS(10,10) in 12 additional $\pi$ orbitals (case d) brings the double bond in excellent 
agreement with the CCSD(T)/CBS value and somewhat shortens the single bond, lowering the BLA to a final value 
of 0.119 \AA. In summary, all choices of multi-determinant expansion in the Jastrow-Slater wave function represent 
a clear improvement with respect to the use of a single determinant, significantly lowering the value of BLA. 
Consequently, the agreement reported in Ref.~\onlinecite{Barborini2015a} between the single-determinant and JAGP 
wave functions indicates that the JAGP ansatz does not have the needed variational flexibility to capture the 
subtle static correlation effects in butadiene.

\begin{figure}[h]
\includegraphics[width=\columnwidth]{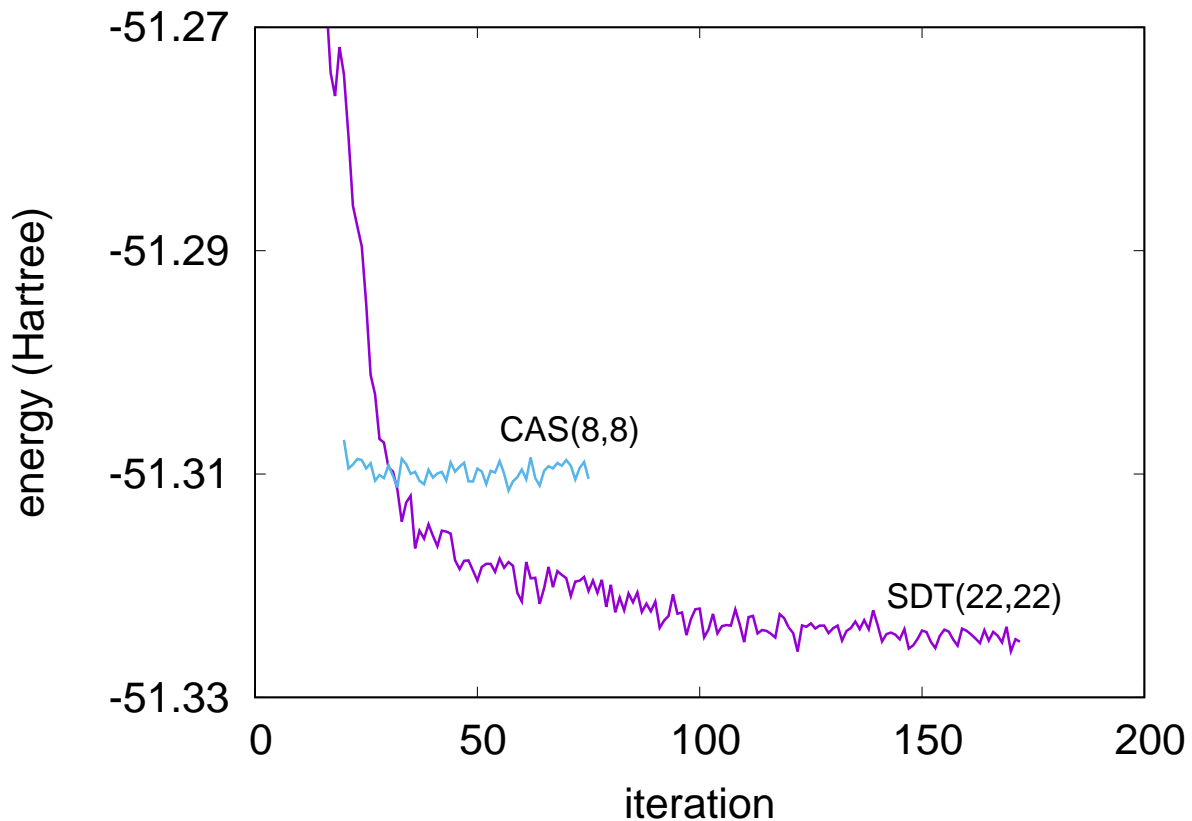}
\caption{
Total energy for a VMC geometry optimization of C$_8$H$_{10}$ using two different expansions in the Jastrow-Slater 
wave function, that is, a full CAS(8,8) with 2468 determinants, and all single, double, and triple excitations in an 
expansion correlating 22 electrons in 22 orbitals for a total of 201924 determinants. The atomic positions and all 
parameters of the wave function (expansion coefficients, orbital and Jastrow parameters) are simultaneously optimized.
}
\label{e_octa}
\end{figure}

Finally, in Fig.~\ref{e_octa}, we demonstrate the ability of our method to optimize the structure and the 
many wave function parameters for the larger molecule C$_8$H$_{10}$ when using a very large determinantal expansion. 
For this purpose, we employ the simple cc-pVDZ basis set and consider all single, double, and triple excitations 
in an expansion denoted as SDT(22,22), correlating 22 electrons in the 22 $\sigma$ 
and $\pi$ orbitals obtained from the carbon valence atomic orbitals. The wave function comprises a total of 201924 
determinants and 58652 parameters. To illustrate the dependence of the energy on the choice of wave function, we 
also display the energy of the last iterations of a structural optimization of the same molecule with the minimal 
CAS(8,8) expansion over the $\pi$ orbitals.  At each iteration, we update both the wave function parameters and 
the atomic positions, the former with one step of the stochastic reconfiguration method and the latter along 
the down-hill direction of the interatomic forces. The energy of the SDT(22,22) wave function is distinctly 
lower than the one obtained with the smaller active space and converged to better than 2 mHartree within 
about 80 iterations. The structural parameters converge much faster and reach stable values
within the first 30 iterations. 

\appendix

\section{Efficient calculation of $Z$, $Y$, $\DI$}

We demonstrate here that we do not need to compute explicitly the inverses of the submatrices $\alpha_I$ as in
Eqs.~(\ref{chi}, \ref{dchi}, and \ref{d2chi}) or in Refs.~\onlinecite{clark11,Filippi_ci2016} to obtain $\DI$ and
its derivatives. These can be computed efficiently using recursion formulas. 

Suppose that $\DI$ contains only third-order excitations (the generalization 
to an arbitrary order  is  straightforward).  Let us rewrite the expression of $\chi$ (Eq.~\ref{chi}) as
\begin{equation}
\DI = \sum _{i_1 < i_2 < i_3,  j_1 < j_2 < j_3} C_{i_1 i_2 i_3 j_1 j_2 j_3}
 \sum_p (-1)^p T_{i_1 p(j_1)} T_{i_2 p(j_2)} T_{i_3 p(j_3)}\,,
\label{chi2}
\end{equation}
where $p$  stands for a permutation of the indices $(j_1, j_2, j_3)$, and $(-1)^p$ is the sign of the permutation.
We note that this formula can also include first- and second-order excitations: a second-order    
excitation  $(i_1 \to j_1, i_2 \to j_2)$ can be written as  $(i_1, \to j_2, i_2 \to j_2, i_3 \to i_3)$, and a 
first-order excitation  $(i_1 \to j_1)$ as $(i_1, \to j_1, i_2 \to i_2, i_3 \to i_3)$.

The starting point is that the tensor of second derivatives can be computed directly from the expression (\ref{chi2}) as
\begin{equation}
\frac{\partial^2 \DI}{\partial T_{i_1 j_1} \partial T_{i_2 j_2}} = \sum_{i_3 j_3} (-1)^{p+q}C_{p(i_1) q(j_1) p(i_2) q(j_2) p(i_3) q(j_3)} T_{i_3 j_3}\,,
\label{d2chi2}
\end{equation}
where $p$ and $q$ are the permutations ordering $(i_1,i_2,i_3)$ and $(j_1,j_2,j_3)$, respectively.
Note that this tensor is antisymmetric with respect to the permutations of either the indices $(i_1,i_2)$ or the 
indices $(j_1,j_2)$, and we only need to compute and store the elements such that $i_1 < i_2$, and $j_1< j_2$.
The tensor of first order derivatives is
\begin{equation}
\frac{\partial \DI}{\partial T_{i_1 j_1}}= \frac{1}{2} \sum_{i_2  j_2} 
\frac{\partial^2 \DI}{\partial T_{i_1 j_1} \partial T_{i_2 j_2}} T_{i_2 j_2}\,,
\label{d1chi2}
\end{equation}
and the value of $\DI$ is 
\begin{equation}
\DI = \frac{1}{3} \sum_{i j}  \frac{\partial \DI}{\partial T_{i j}} T_{ij} \,. 
\label{d0chi2}
\end{equation}
In practice, sparse representations of these tensors should be used.
The formula (\ref{d2chi2}) involves at most nine products and nine sums per excitation. The formulas 
(\ref{d1chi2}) and  (\ref{d0chi2}) require less than $N^2 N^2_{\rm orb}$   and  $N N_{\rm orb}$  operations 
(additions  or multiplications), respectively.
The method still scales like $O(N_e)$ but with a reduced  prefactor because no divisions are involved and the 
number of operations is smaller.  For example, expression (\ref{d0chi2}) involves at most $NN_{\rm orb}$ multiplications 
and additions  whereas (\ref{chi}) is a sum on $N_e$ terms ($N_e$ can be of order $N^3N^3_{\rm orb}$ if third-order 
excitations are included). 


\section{One-electron-move algorithms}

To sample the density $\Psi^2$, we use the Metropolis-Hastings method \cite{hastings70,TOULOUSE2016285} which is 
a stochastic dynamics  in the space of configurations  ${\bf R}=  ({\bf r_1}, {\bf r}_2 \dots {\bf r}_N)$.
For a given iteration, this method proposes a random move  ${\bf R} \to {\bf R}^\prime$ with a transition 
probability density $P({\bf R} \to {\bf R}^\prime)$.  The proposed move is accepted with the probability 
\begin{equation}
{\rm min} \left(\frac{\Psi^2 ({\bf R}^\prime)}{\Psi^2 ({\bf R})} \frac{P({\bf R}^\prime \to {\bf R})}{P({\bf R} \to {\bf R}^\prime)}, 1 \right)\,.
\end{equation}
If only one electron is moved (here the first, for example),  the new configuration is 
${\bf R}^\prime =  ({\bf r_1}^\prime , {\bf r}_2 \dots {\bf r}_N)$.
The new extended Slater matrix $\tilde{A}^\prime$ differs from $\tilde{A}$ only in the first line. 

We introduce the matrix $\tilde{B}_e$ such that the first line of $\tilde{B}_e$ and $\tilde{A}^\prime$  are the same
but  $\tilde{B}_e$ is zero elsewhere.  Since $\Phi$ is a linear function of the modified line
\begin{eqnarray}
\frac{  \Phi{(\bf R}^\prime)}{\Phi ({\bf R})} &=& \partial_\lambda \ln \Phi(\tilde{A})\,,
\end{eqnarray}
where we considered the following transformation $\tilde{A} \to \tilde{A}+\lambda \tilde{B}_e$.
Using  Eq.~(\ref{ymder}), we obtain
\begin{eqnarray}
\frac{ \Phi({\bf R}^\prime)}{\Phi ({\bf R})} =
  \tr(A^{-1}  \partial_\lambda A ) + \tr(Y  \partial_\lambda \T  )\,,
\label{ya}
\end{eqnarray}
where we recall that $\T= A^{-1}\tilde{A}$ and $\partial_\lambda \T = A^{-1}\tilde{B}_e -A^{-1}B_e \T $.
The cost of this calculation is $O(NN_{\rm orb}) \sim O(N^2)$. 
When the first electron has been moved, $\T$ can be updated using the Sherman Morrison formula at a cost 
$O(NN_{\rm orb})$\cite{Filippi_ci2016}, and  $Y$ which depends on $\T$ can  be again computed at a cost 
$O(N_e)$.  The  total cost for a sweep (each electron has moved once) is  $O(N^2N_{\rm orb}) + O(NN_e)$.
The matrix $\Gamma$ and all derivatives are computed  after each sweep. 

We note that, if one uses instead the expression involving $\Gamma$ to update the wave function,
\begin{equation}
\frac{ \Phi({\bf R}^\prime)}{\Phi ({\bf R})}= \tr(\Gamma \tilde{B}_e)\,,
\label{gammae}
\end{equation}
one would need to update $\Gamma$ at each Monte Carlo step and incur the higher cost of $O(N^4) + O(N N_e)$ for a full sweep. 
This is because  updating $\Gamma$ requires the calculation of  $\partial_\lambda \Gamma$ given in  Eq.~(\ref{gammader}), 
where of course $\tilde{B}$ is replaced by $\tilde{B}_e$. In this equation, the product $(\partial_\lambda Y)A^{-1}$ 
scales like  $O(N^3)$, unless $Y$ is sparsely modified after  one electron move (i.e. a few double excitations are involved).  

Finally, also in the calculation of the drift of a single electron $\nabla_i \Phi/\Phi$ needed in the 
Monte Carlo sampling, it is better not to recompute $\Gamma$ but to use formula (\ref{ya}) with 
$\partial_\lambda \T = A^{-1}\tilde{B}^{\rm drift}_e -A^{-1}B^{\rm drift}_e \T $, where the matrix 
$\tilde{B}^{\rm drift}_e$ is zero except the $i^{th}$ row which equals $\nabla \phi_j ({\bf r}_i)$.
However, if the sampling is modified to use a finite distribution at the nodes following 
Ref.~\onlinecite{Attaccalite2008}, the full drift has to be computed at each step.  The resulting 
scaling is  $O(N^4)+O(N N_e)$ per sweep, using Eq.~(\ref{ya}) or (\ref{gammae}) alike.

\section{Simple Expression  of $\Gamma$ for a Jastrow-Slater expansion}

Here, we provide a simple (though not  efficient) expression for $\Gamma$ and some mathematical properties.

\subsection*{Simple expression for $\Gamma$}

The determinantal contribution of the wave function written in  Eq.~(\ref{phidef}) is 
\begin{equation}
\Phi =  \sum_{I=0}^{N_e} c_I \det ({A}_I) \,. 
\nonumber
\end{equation}
where $A_I$ is a list of $N$ columns of the $N \times N_{\rm orb}$ generalized Slater matrix  $\tilde{A}$.  
We  can then  define a $N_{\rm orb}\times N$ matrix $R_I$ such that 
\begin{equation}
A_I = \tilde{A} R_I'\,,
\end{equation}
which gives an explicit expression of $\Phi$ as a function of $\tilde{A}$
\begin{equation}
\Phi (\tilde{A})= \sum_I c_I \det(\tilde{A}R_I) \,.
\end{equation}
For example, given a $3\times 3$ Slater matrix built on the orbitals $(\phi_1,\phi_3, \phi_4)$
\begin{equation}
R_I = \left( \begin{array}{cccccc}
1 & 0  & 0   \\ 
0 & 0  & 0  \\ 
0 & 1  & 0  \\
0 & 0  & 1 \\
0 & 0 & 0 \\
\vdots & \vdots \\
\end{array} \right)\,.
\nonumber
\end{equation} 
The derivative of the determinantal expansion with respect to a parameter $\mu$ is 
\begin{eqnarray}
\partial_\mu \Phi & =&  \sum_{I} c_I \det (A_I) \tr(A_I^{-1}  \partial_\mu A_I) \nonumber \\
 & = &  \sum_{I} c_I \det (A_I) \tr(A_I^{-1}  \partial_\mu \tilde{A} R_I )\,. \nonumber
\end{eqnarray}
Using the linearity and the cyclic properties of the trace, we find  
\begin{equation}
\frac{\partial_\mu \Phi}{\Phi} =  \tr(\Gamma \partial_\mu \tilde{A})\,,
\label{gentrace}
\end{equation}
where we can identify $\Gamma$
\begin{equation}
\Gamma =    \frac{1}{\Phi} \sum_{I} c_I \det (A_I) R_I A_I^{-1} \,.
\label{gamma_e2}
\end{equation}
In the expression (\ref{gamma_e2}), the application of  $R_I$ on the left of $A_I^{-1}$ dispatches 
the $N$ lines of $A_I^{-1}$ in a  larger $N_{\rm orb}\times N$  matrix.  Of course, a direct evaluation 
of (\ref{gamma_e2}) would be  $O(N_e N^3)$ and would be too costly. 

\subsection*{Properties of the matrix $\Gamma$}

$\Gamma$ is a right inverse of $\tilde{A}$, i.e. 
\begin{equation}
\tilde{A}\Gamma =  I_N\,,
\label{cofactgen}
\end{equation}
where $I_N$ is the identity matrix of order $N$.
The proof is simple 
\begin{eqnarray}
\tilde{A} \Gamma &=& \frac{1}{\Phi}  \sum_I c_I \det(A_I) \tilde{A} R_I A_I^{-1} \\
 & = & \frac{1}{\Phi} \sum_I c_I \det(A_I) A_I A_I^{-1} =   I_N  \,.
\end{eqnarray}
We now consider the $N_{\rm orb}\times N_{\rm orb}$ matrix $\Gamma \tilde{A}$ and
resort to the transformation $\phi_i \to \phi_i + \mu_{ij} \phi_j$.  The only non-zero column of the matrix 
${\partial \tilde{A}}/{\partial \mu_{ij}}$   is the $i^{th}$ column, which is the same as  the  $j^{th}$  column of  $\tilde{A}$.  
Therefore,
\begin{equation}
\frac{1}{\Phi}\frac{\partial \Phi}{\partial \mu_{ij}} = \tr\left(\Gamma \frac{\partial \tilde{A}}{\partial \mu_{ij}}\right) 
= (\Gamma \tilde{A})_{ij}\,,
\end{equation}
meaning that  $\Phi(\Gamma \tilde{A})_{ij}$ is  the new value of the determinantal expansion when the orbital $i$ has been 
replaced by the orbital $j$
\begin{equation}
\Phi(\Gamma \tilde{A})_{ij} = \sum_{I} c_I \det(A_I^{i \to j})\,.
\end{equation}
In particular, if $i=j$, 
\begin{equation}
\Phi(\Gamma \tilde{A})_{ii} = \sum_{I / \phi_i \in A_I} c_I \det(A_I)\,.
\end{equation}
In other words, the main diagonal of $\Phi \Gamma \tilde{A}$ is made of restrictions of the summation in 
(\ref{phidef}) to  determinants containing a given orbital.
As a by-product, if $\phi_i$ is common to all the determinants of the expansion, $ (\Gamma \tilde{A})_{ii}$ is equal to  $1$.  
If $i\ne j$, $\Phi(\Gamma \tilde{A})_{ij}$ is the expansion (\ref{phidef}) restricted to Slater determinants  occupied by $\phi_i$ and not  by $\phi_j$
 \begin{equation}
\Phi(\Gamma \tilde{A})_{ij} = \sum_{I / \phi_i \in A_I, \Phi_j \not\in A_I} c_I \det(A_I^{i\to j}) \,.
\end{equation}
In particular, if the orbital $j$ is common to all determinants, $(\Gamma \tilde{A})_{ij}=0$ for any $i\ne 0$. 
In conclusion, if there are $N_{\rm act}$ orbitals which can be excited (i.e.\ there are $N-N_{\rm act}$ orbitals common 
to all determinants), the following property holds:  $\Gamma \tilde{A}$ contains a $N_{\rm orb}\times (N-N_{\rm act})$ 
block which is zero with the  exception of  a $(N-N_{\rm act})\times (N-N_{\rm act})$ square sub-block which is the 
identity matrix. 


\section{Calculation of $\Gamma$ using the Sherman-Morrison-Woodbury formula}

Here, we  derive  the expression (\ref{gammeff}) directly from the identity (\ref{gamma_e2}) using 
the Sherman-Morrsion-Woodbury formula.  The algebra is a bit more tedious. First, we remind some notations 
useful to explicit the matrix $R_I$ and  dependencies on $\tilde{A}$.  $A$ is the reference Slater matrix 
and $\P$ is the matrix  which selects the columns  $\tilde{A}$ from which  $A$ is made
\begin{equation}
A=\tilde{A}\P \,.
\end{equation}
$P_I$ is the matrix  such that $A P_I$ is the list of the $k_I$ columns of $A$ which differ from those of 
$A_I$ (see for example Eq.~(\ref{expi})).  The $N \times N$ matrix  $P_I P_I^T$ is a diagonal matrix: 
if $i$ is the index of a column which differ in $A$ and $A_I$, $(P_I P_I^T)_{ii}=1$, while $(P_I P_I^T)_{ii}=0$ 
otherwise.  Consequently, the identity  
\begin{equation}
A_I-A = (A_I-A) P_I  P_I^T
\end{equation}
holds.
The list of excited orbitals are the columns of $A_I P_I$ and can be selected from $\tilde{A}$ with the aid of the
$N_{\rm orb}\times k_I$ matrix $Q_I$  such that 
\begin{equation}
A_I P_I = \tilde{A} Q_I\,,
\end{equation}
as in the example Eq.~(\ref{exqi}).  With these definitions 
\begin{equation}
A_I =  A+(A_I-A) P_I P_I^T = \tilde{A} \left(\P+(Q_I-\P P_I)P_I^T     \right) \,,
\end{equation}
and the matrix $R_I$ which selects the columns  $\tilde{A}$ from which  $A_I$, is given by
\begin{equation}
R_I = \P+(Q_I-\P P_I)P_I^T\,.
\end{equation}
Now, writing $A_I=A + (A_I-A)P_I P_I^T$ and applying the Sherman-Morrison-Woodbury formula, we obtain
\begin{eqnarray}
A_I^{-1} & =& A^{-1} -A^{-1} (A_I-A)P_I (1+ P_I^T A^{-1} (A_I-A)P_I)^{-1} \, P_I^T A^{-1} \nonumber \\
 & =&  A^{-1}-A^{-1} (A_I-A)P_I (P_I^T A^{-1} A_I P_I)^{-1} \, \, P_I^T A^{-1}\,,
\label{SMW} 
\end{eqnarray}
so that
\begin{equation}
A_I^{-1} = A^{-1}+P_I\alpha_I^{-1} P_I^{T}A^{-1} -A^{-1}\tilde{A}Q_I \alpha_I^{-1}P_I^{T}A^{-1} \,,
\label{SMW2}
\end{equation}
where we have introduced
\begin{equation}
\alpha_I \equiv  P_I^T A^{-1}\tilde{A} Q_I\,.
\end{equation}
Multiplying both sides of Eq.~(\ref{SMW2}) by $P_I^T$ gives the  following identity 
\begin{eqnarray}
P_I^T A_I^{-1} & =& P_I^T P_I\alpha_I^{-1} \, \, P_I^T A^{-1} \nonumber \\
 & =& \alpha_I^{-1} \, \, P_I^T A^{-1}\,.
\label{pbara}
\end{eqnarray}
Using this expression, we can simplify 
\begin{eqnarray}
R_I A_I^{-1}&=& \P A_I^{-1}+(Q_I-\P \P_I)\alpha_I^{-1} \, \, P_I^T A^{-1}  \nonumber \\
 &= & \P A^{-1} -\P A^{-1}\tilde{A}Q_I \alpha_I^{-1} P_I^T A^{-1}+Q_I \alpha_I^{-1}P^T_I A^{-1} \nonumber \\
& = & \P A^{-1} +(1-\P A^{-1}\tilde{A})Q_I\alpha_I^{-1}P_I^{T} A^{-1}
\label{riai}
\end{eqnarray}
From equations (\ref{riai}) and  (\ref{gamma_e2}), we then obtain
\begin{equation}
\Gamma  =   \P A^{-1} +\ (1-\P A^{-1}\tilde{A}) Y A^{-1}\,,
\label{gammacompact}
\end{equation}
with
\begin{equation}
Y \equiv \frac{\det(A)}{\Phi} \sum_I c_I \det(\alpha_I) Q_I \alpha_I^{-1} P_I^T
\label{defY}
\end{equation}
and, of course,
\begin{eqnarray}
  \Phi & = & \det(A)\left(\sum_{I=1}^{N_e} c_I \det(\alpha_I) \right) \,. 
\end{eqnarray}

\acknowledgments
C.F.\ acknowledges support from the Netherlands Organization for Scientific Research (NWO) for the use of 
the SURFsara supercomputer facilities.

\bibliography{cipaper}

\begin{thebibliography}{21}
\bibitem [{Hps()}]{Hpseudo}%
  \BibitemOpen
  \href@noop {} {}\bibinfo {note} {For the hydrogen atom, we use a more
  accurate BFD pseudopotential and basis set. Dolg, M.; Filippi, C., private
  communication}\BibitemShut {NoStop}%
\end{thebibliography}
\end{document}